\documentclass[12pt]{article}
\usepackage[margin=1in]{geometry} 
\usepackage{setspace}
\usepackage{placeins}
\usepackage{etoolbox}
\usepackage{natbib}
\usepackage{url} 
\usepackage{bbm} 
\usepackage{dsfont}
\usepackage{graphicx}%
\usepackage{multirow}%
\usepackage{amsmath,amssymb,amsfonts}%
\usepackage{amsthm}%
\usepackage[title]{appendix}%
\usepackage{xcolor}%
\usepackage{textcomp}%
\usepackage{manyfoot}%
\usepackage{booktabs}%
\usepackage{algorithm}%
\usepackage{algorithmicx}%
\usepackage{algpseudocode}%
\usepackage{listings}%
\usepackage{tikz}
\usetikzlibrary{shapes.geometric, arrows}
\usetikzlibrary{arrows.meta}
\usetikzlibrary{shapes,backgrounds,calc,positioning}
\usepackage{array}
\usepackage{blkarray}
\usepackage{comment}
\usepackage{booktabs}

\newcommand{\E}{\mathbb{E}}

\newcommand{\Var}{\operatorname{Var}}
\newcommand{\V}{\operatorname{Var}}

\newcommand{\I}{\mathds{1}}

\newcommand{\bD}{\mathbf{D}}
\newcommand{\bU}{\mathbf{U}}
\newcommand{\bZ}{\mathbf{Z}}
\newcommand{\balpha}{\boldsymbol{\alpha}}
\newcommand{\bbeta}{\boldsymbol{\beta}}
\newcommand{\btheta}{\boldsymbol{\theta}}
\newcommand{\bv}{\boldsymbol{v}}
\newcommand{\bA}{\mathbf{A}}
\newcommand{\bB}{\mathbf{B}}
\newcommand{\bI}{\mathbf{I}}

\newtheorem{theorem}{Theorem}

\newtheorem{lemma}{Lemma}

\title{Two-Stage Least Squares Instrumental Variable Estimation for Semiparametric Accelerated Failure Time Models with Right-Censored Data}

\author{Zian Zhuang$^{1}$,
Hua Zhou$^{1}$, 
Jin Zhou$^{1}$, and
Gang Li$^{1^*}$\\
$^{1}$Department of Biostatistics, University of California at Los Angeles \\
$^*$ Corresponding author(s).
}

\begin{document}
\FloatBarrier        

\clearpage 



\maketitle
\label{firstpage}


\begin{abstract}
Instrumental variable (IV) analysis is widely used in fields such as economics and epidemiology to address unobserved confounding and measurement error when estimating the causal effects of intermediate covariates on outcomes. However, extending the commonly used two-stage least squares (TSLS) approach to survival settings is nontrivial due to censoring. This paper introduces a novel extension of TSLS to the semiparametric accelerated failure time (AFT) model with right-censored data, supported by rigorous theoretical justification. Specifically, we propose an iterative reweighted generalized estimating equation (GEE) approach that incorporates Leurgans’ synthetic variable method, establish the asymptotic properties of the resulting estimator, and derive a consistent variance estimator, enabling valid causal inference. Simulation studies are conducted to evaluate the finite-sample performance of the proposed method across different scenarios. The results show that it outperforms the naïve unweighted GEE method, a parametric IV approach, and a one-stage estimator without IV. The proposed method is also highly scalable to large datasets, achieving a 300- to 1500-fold speedup relative to a Bayesian parametric IV approach in both simulations and the real-data example.  We further illustrate the utility of the proposed method through a real-data application using the UK Biobank data.

\end{abstract}

%

\textbf{keywords}: synthetic variables, two-stage least squares (TSLS), censored two-stage least squares (cTSLS), censored data, instrumental variable (IV) analysis.

\section{Introduction}\label{sec1}

Causal inference with observational data is often complicated by unobserved confounding and measurement error in covariates. Instrumental variable (IV) methods provide a principled framework to address these challenges and have a long history in econometrics and statistics \citep{wright1921correlation, wright1928tariff, haavelmo1943statistical, haavelmo1944probability, theil1958economic}. The two-stage least squares (TSLS) approach, in particular, has become the canonical tool for consistent estimation under endogeneity \citep{goldberger1972structural, heckman1985alternative, heckman1989choosing} and has since been widely applied across disciplines. In epidemiology, Mendelian randomization has further popularized IV analysis by using genetic variants as instruments to strengthen causal inference \citep{gray1991avoid, davey2003mendelian, smith2004mendelian, didelez2007mendelian, lawlor2008mendelian}.

Extending IV methods to time-to-event outcomes introduces additional challenges due to censoring. Several IV methods have been proposed for the Cox model, including score-based estimators with additive confounding terms \citep{mackenzie2014using}, two-stage residual inclusion (2SRI) frailty procedures \citep{martinez2019adjusting}, structural estimators \citep{martinussen2019instrumental_nocolla, sorensen2019causal_nocolla}, and inverse probability weighting approaches \citep{kianian2021causal}. These methods, however, face difficulties: IV assumptions may conflict with the proportional hazards model, and 2SLS/2SRI estimators are susceptible to non-collapsibility bias \citep{cai2011two, hernan2010hazards, li2015instrumental, wan2015bias, wan2018general, dukes2019doubly}. To overcome these limitations, IV methods have been extended to Aalen’s additive hazards model \citep{martinussen2017instrumental, tchetgen2015instrumental, li2015instrumental, brueckner2019instrumental, dukes2019doubly, zhao2025additive}, the generalized accelerated failure time (GAFT) model \citep{bijwaard2009instrumental}, competing risks settings \citep{kjaersgaard2016instrumental, richardson2017nonparametric, zheng2017instrumental, martinussen2020instrumental, beyhum2023nonparametric}, and interval-censored survival data through nonparametric, doubly robust, and machine learning–based approaches \citep{li2023instrumental, lou2025instrumental, lee2023doubly, wang2023instrumental_nocolla, crommen2025estimation, junwen2024doubly, cheng2024instrumental}.

Another line of research has developed IV approaches for accelerated failure time (AFT) models, including parametric Bayesian formulations \citep{li2015bayesian, crommen2024instrumental} and semiparametric linear rank procedures \citep{huling2019instrumental}. IV methods based on AFT models are particularly appealing because they provide interpretable covariate effects on survival times and align naturally with classical IV frameworks such as TSLS. Nonetheless, existing AFT-based IV methods have important limitations. Parametric approaches rely on restrictive distributional assumptions and are highly sensitive to model misspecification, as illustrated in our simulations and empirical analysis. The semiparametric linear rank method of \citet{huling2019instrumental} is restricted to a single binary instrument, and although extensions to multiple or continuous instruments are theoretically feasible, they entail substantial additional complexity; moreover, the associated estimating equations may be non-monotone, even with Gehan weights, leading to multiple roots and numerical instabilities. Taken together, these limitations highlight the need for IV methods for AFT models that are robust to distributional misspecification, computationally stable, and applicable with a broader class of instruments.

This paper develops a new IV method for the semiparametric AFT model that incorporates the synthetic variable approach for censored linear regression \citep{koul1981regression, Leurgans87RandomCensoring, Zheng87PseudoVariableCensor, lai1995asymptotic}. The proposed framework avoids parametric distributional assumptions, accommodates a broader class of instruments, and mitigates the numerical instability inherent in existing rank-based procedures. We extend the TSLS framework to the semiparametric AFT setting with right-censored data by formulating a reweighted generalized estimating equation (GEE) estimator that incorporates Leurgans’ synthetic variable method to account for censoring. We refer to the resulting estimator as the censored TSLS (cTSLS) estimator.
 The large-sample properties of the proposed estimator are established by extending the asymptotic theory of \citet{lai1995asymptotic} to the IV–AFT framework, and a consistent sandwich variance estimator is derived to facilitate valid inference.

Finite-sample performance is evaluated through simulation studies comparing bias, variance, and coverage probability. The results demonstrate that the proposed cTSLS method possesses superior finite-sample properties, outperforming the naïve unweighted GEE estimator, the censored parametric Bayesian instrumental variable analysis approach (PBIV) \citep{li2015bayesian}, and a one–stage censored least–squares estimator using Leurgans’ synthetic outcomes without IV adjustment (cOLS). Across the scenarios considered, cTSLS yields unbiased estimates with nominal coverage, outperforms the other models, and is dramatically faster—more than 300× faster than PBIV at comparable sample sizes. Finally, we illustrate the practical utility of the proposed method in a real-data application using the UK Biobank to assess the causal effect of systolic blood pressure (SBP) on time to cardiovascular disease (CVD) diagnosis among individuals with diabetes.

The remainder of this paper is organized as follows. 
Section~\ref{sec:method} begins by reviewing the classical TSLS method and then discusses the shortcomings of the naive extension of TSLS with synthetic variables under censoring. 
We then introduce the proposed cTSLS estimator for the semiparametric IV-AFT model, including the synthetic variable construction, the iterative weighting scheme, and the asymptotic properties of the estimator. 
Section~\ref{sec:sim_study} reports the results of simulation studies evaluating bias, variance estimation, coverage probability, and computation time. 
Section~\ref{sec:real_study} illustrates the application of the method to the UK Biobank data and compares it with alternative approaches.  
Finally, Section~\ref{sec:discussion} concludes with a discussion of the findings and directions for future research.

\section{Method}\label{sec:method}

\subsection{Notation and preliminaries} \label{sec:notation}
Causal inference is commonly formulated within two complementary frameworks: the potential outcomes framework \citep{neyman1923applications, rubin1974estimating, imbens2015causal, HernanRobins2020} and the causal diagram framework \citep{pearl2009causality}, which are mathematically connected \citep{richardson2013single}. In this article, we adopt the causal diagram framework and use directed acyclic graphs (DAGs) together with structural equations to formally state identification assumptions and derive the corresponding estimands. We begin by introducing the notation, structural equations, and assumptions underlying the DAG-based formulation.

\subsubsection{DAG-Based Formulation and Assumptions}
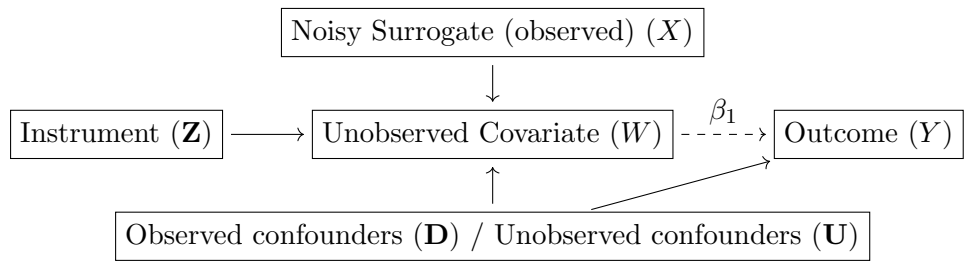
\begin{figure}[ht]

\begin{tikzpicture}[node distance=5cm]
  \small
  \node (instrument) [rectangle, draw] {Instrument ($\bZ$)} ; 
  \node (unobserved) [rectangle, draw] [right of=instrument] {Unobserved Covariate ($W$)};
  \node (outcome) [rectangle, draw, right of=unobserved] {Outcome ($Y$)};
  \node (Noisy) [rectangle, draw, above=0.7cm of unobserved] {Noisy Surrogate (observed) ($X$)};
  \node (confounds) [rectangle, draw, below=0.7cm of unobserved, align=center] {Observed confounders ($\bD$) / Unobserved confounders ($\bU$)};
  \draw [->, shorten >=3pt, shorten <=3pt] (instrument) -- (unobserved);
  \draw [->, dashed, shorten >=3pt, shorten <=3pt] (unobserved) -- node [pos=0.5, sloped, above] {$\beta_1$} (outcome);
  \draw [->, shorten >=3pt, shorten <=3pt] (Noisy) -- (unobserved);
  \draw [->, shorten >=3pt, shorten <=3pt] (confounds) -- (unobserved);
  \draw [->, shorten >=3pt, shorten <=3pt] (confounds) -- (outcome);
\end{tikzpicture}

\caption{Directed acyclic graph (DAG) for instrumental variable analysis. 
}
\label{fig:dag-iv-model}
\end{figure}

Figure~\ref{fig:dag-iv-model} illustrates the IV DAG considered in this paper, where
$Y$ denotes the outcome, $W$ the unobserved endogenous covariate, $X$ its observed surrogate, $\bZ$ the vector of instrumental variables, and $\bD$ and $\bU$ the observed and unobserved confounders. 
The parameter $\beta_1$ represents the causal effect of $W$ on $Y$. 
Directed arrows indicate causal relationships. 

Assume that the instrument $\bZ$ satisfies the standard IV conditions: (i) \emph{relevance} ($\bZ\mathrel{\not\!\perp\!\!\!\perp} W$), (ii) \emph{independence} ($\bZ\mathrel{\perp\!\!\!\perp}\bU|\bD$ and independent of measurement error in $W$), and (iii) \emph{exclusion} ($\bZ\mathrel{\perp\!\!\!\perp} Y|W,\bD$). 

With these assumptions in place, we formalize the IV model for the semiparametric accelerated failure time (IV–AFT) framework under the DAG representation:
\begin{align}
W_i &= \alpha_0 + \balpha_1^{\top}\bZ_i + \balpha_2^{\top}\bD_i + \balpha_3^{\top}\bU_i + \tilde\epsilon_{1i}, \label{eq:W} \\
Y_i &= \beta_0 + \beta_1 W_i + \bbeta_2^{\top}\bD_i + \bbeta_3^{\top}\bU_i + \tilde\epsilon_{2i}, \label{eq:Y}\\
X_i &= W_i + \tilde\epsilon_{3i}, \label{eq:XW}
\end{align}
where $Y_i$ denotes the log-transformed event time
for $i = 1, \dots, n$. The errors $\tilde\epsilon_{1i}, \tilde\epsilon_{2i},$ and $\tilde\epsilon_{3i}$ are assumed to have mean zero and finite variances $\tau_1^2, \tau_2^2,$ and $\tau_3^2$, respectively. Furthermore, $\tilde\epsilon_{1i}, \tilde\epsilon_{2i}, \tilde\epsilon_{3i}, \bU_i$, and $\bZ_i$ are assumed to be mutually independent. Without loss of generality, we assume $\bU_i$ has mean zero and finite variances.
The coefficient $\beta_1$ represents the endogenous effect of primary interest.

Since $W_i$ is not directly observed, we replace $W_i$ with $X_i - \tilde\epsilon_{3i}$ in Equations~(\ref{eq:W}) and~(\ref{eq:Y}). Under this substitution, the semiparametric IV–AFT model \eqref{eq:W}--\eqref{eq:XW} is equivalently expressed as:
\begin{align}
    X_i &= \alpha_0 + \balpha_1^{\top} \bZ_i + \balpha_2^{\top}\bD_i + \epsilon_{1i}, \label{eq:X_i}\\
    Y_i &= \beta_0 + \beta_1 X_i + \bbeta_2^{\top}\bD_i + \epsilon_{2i}, \label{eq:Y_i}
\end{align}
where  $
\epsilon_{1i} := \balpha_3^{\top}\bU_i + \tilde\epsilon_{1i} + \tilde\epsilon_{3i}, 
$ and
$\epsilon_{2i} :=  \bbeta_3^{\top}\bU_i + \tilde\epsilon_{2i} - \beta_1 \tilde\epsilon_{3i}$
represent the composite errors.

Since $\epsilon_{2i}$ and $(X_i, \bD_i)$ may be correlated through the unobserved confounders $\bU_i$, directly regressing $Y$ on $(X_i, \bD_i)$ using least squares yields a biased and generally inconsistent estimator of $\beta_1$. 
To address this, we review the classical two-stage least squares (TSLS) method for consistently estimating $\beta_1$ in the absence of censoring.

\subsubsection{The Classical Two-Stage Least Squares (TSLS) Method}\label{sec:tsls}

The classical two-stage least squares (TSLS) method 
\citep{theil1958economic, goldberger1972structural, imbens2015causal} proceeds in two steps. 
In the first stage, $X_i$ is regressed on instruments $\bZ_i$ and observed covariates $\bD_i$ using ordinary least squares (OLS), yielding fitted values  
\[
\hat X_{i} = \hat\alpha_{0} + \hat\balpha_{1}^{\top}\bZ_{i} + \hat\balpha_{2}^{\top}\bD_i.
\]  
In the second stage, $Y_i$ is regressed on $\hat X_i$ and $\bD_i$ using OLS, producing consistent estimators $\hat\beta_{0}, \hat\beta_{1},$ and $\hat\bbeta_{2}$.  

For inference on the endogenous effect parameter $\beta_1$, the variance estimator of $\hat\beta_1$ must account for the sampling variability introduced in the first stage.
This can be derived by considering the stacked parameter vector $\btheta = (\balpha^\top, \bbeta^\top)^\top$. Since the TSLS estimator $\hat\btheta$ can be viewed as the solution to estimating equations arising from the two regressions, and its asymptotic variance-covariance matrix is consistently estimated by the robust sandwich form 
\citep{white1982maximum, hansen1982large, newey1994large, wooldridge2010econometric}
\begin{equation}
\label{eq:sandwich0}
\widehat{\Var}(\hat\btheta) = \tilde\bA^{-1} \tilde\bB \tilde\bA^{-{\top}}, 
\end{equation}  
where $\tilde\bA$ and $\tilde\bB$ are sample analogues of the Jacobian and covariance of the estimating functions.  
This robust variance estimator  is a special case of the generalized method of moments (GMM) framework \citep{hansen1982large, newey1994large}.

\subsection{Two-Stage Least Squares Estimation for Right-Censored Data (cTSLS)}
Now we consider estimation and inference for $\beta_1$ based on a right-censored sample data of the form  
\[
\left\{ (\tilde{Y}_i, \delta_i, X_i, \bD_i, \bZ_i), \quad i=1,\ldots, n \right\},
\]
where $\tilde{Y}_i = \min(Y_i, C_i)$ is the observed log follow-up time, $\delta_i = I(Y_i \leq C_i)$ is the censoring indicator, $Y_i$ is the log event time of interest, and $C_i$ is the log censoring time. Here $X_i$ denotes the observed surrogate for $W_i$, $\bD_i$  the vector of observed confounders and $\bZ_i$ the vector of instruments. For simplicity, we assume noninformative right censoring, i.e., $C_i \perp (Y_i, \bD_i, \bZ_i)$, so that the censoring mechanism is independent of the event process as well as the covariates and instruments. Nevertheless, the proposed method can be extended to the weaker assumption $C_i \perp Y_i \mid (\bD_i, \bZ_i)$ by incorporating a regression model for the censoring distribution, as is commonly done in the IPCW literature \citep{robins1992recovery, zeng2007efficient}.

When outcomes are subject to right censoring, the classical TSLS method is infeasible because $Y_i$ is not fully observed. 
To address this challenge, we develop a censored two-stage least squares (cTSLS) method for the semiparametric IV-AFT model \eqref{eq:X_i}-\eqref{eq:Y_i} that incorporates synthetic variables in the second stage. 
The key idea is to construct synthetic outcomes $Y_{i\hat{G}}^*$ that serve as approximately unbiased surrogates for the unobserved $Y_i$. 
Two synthetic variable constructions that have been well studied in censored linear regression were proposed by \citet{koul1981regression} and \citet{Leurgans87RandomCensoring}, with \citet{Zheng87PseudoVariableCensor} showing that the Leurgans formulation yields smaller variance. 
Accordingly, we adopt the Leurgans construction throughout this paper.

Let $G(t) = P(C \le t)$ denote the distribution function of the log censoring time $C$.
For the oracle case with known $G$, the Leurgans synthetic variable is defined as
\begin{align}
    Y_{iG}^* = \tilde{Y}_i + \int_{-\infty}^{\tilde{Y}_i} \frac{G(t-)}{1 - G(t-)}  dt. \label{eq: sv leurgans} 
\end{align}
It can be shown that $\E(Y_{iG}^* \mid Y_i) = Y_i$, which
implies $\E(Y_{iG}^* ) = \E(Y_i)$ and thus justifies it uses in linear regression.
In practice, however, $G$ is unknown and is replaced by the Kaplan--Meier estimator $\hat G$ \citep{kaplan1958nonparametric}, yielding
\begin{align}
    Y_{i\hat G}^* = \tilde{Y}_i + \int_{-\infty}^{\tilde{Y}_i} \frac{\hat G(t-)}{1 - \hat G(t-)}  dt. \label{eq: sv leurgans kme}
\end{align}

\subsubsection{Flaws of the Naive Censored TSLS Method} \label{sec:compare method}

A naive extension of the classical TSLS method to right-censored data is to substitute $Y_{i\hat G}^*$ for $Y_i$ in the TSLS procedure described in Section~\ref{sec:tsls}, for both point estimation and variance estimation. 
This strategy has also been suggested in earlier unpublished work using the synthetic variable of \citet{koul1981regression} \citep[see, e.g.,][]{atiyat2011instrumental, wang2017methods}. 
However, this approach is fundamentally flawed for several reasons.  

First, the synthetic variables $Y_{i\hat G}^*$ are not independent across subjects because they all depend on the estimated $\hat G$. 
This dependence complicates the derivation of asymptotic properties such as consistency and asymptotic normality, and—most importantly—introduces an additional source of variability in the resulting cTSLS estimator. 
Such variability cannot be accommodated by simply substituting $Y_{i\hat G}^*$ for $Y_i$ in the classical sandwich variance estimator \eqref{eq:sandwich0}.  

Second, unlike the original outcome variable $Y_i$, the oracle synthetic variables $Y_{iG}^*$ are inherently heterogeneous, even when $G$ is known, because their variances differ across subjects. 
This heterogeneity renders least squares estimation inefficient.  

To address these issues, we propose an iterative weighted estimator that improves efficiency and rigorously establish its asymptotic normality, with its asymptotic variance consistently estimated by a robust sandwich estimator, as developed in the following subsections.

\subsubsection{cTSLS Estimator}
Our proposed cTSLS estimator extends the classical TSLS procedure by modifying only the second stage. 
Specifically, in place of $Y_i$, we substitute the synthetic outcome $Y_{i\hat G}^*$ 
and employ weighted least squares regression with subject–specific weights 
to account for heterogeneity:
\begin{align}
\omega_i 
= \frac{1}{\Var(Y_{iG}^*)} 
= \frac{1}{\Var(Y_i) 
   + 2 \int_{G^{-1}(0)}^\infty \left\{1 - F_i(s)\right\} 
      \int_{-\infty}^{\mu_{Y_i}+s} \frac{G(t)}{1-G(t)}  dt  ds}, 
   \quad i=1,\ldots,n, \label{eqn:sv_var}
\end{align}
where $F_i(s) = P(\epsilon_{2i} \leq s)$ denotes the distribution of the second–stage error. 
A detailed derivation of $\Var(Y_{i G}^*)$ is provided in Appendix~\ref{section:var_sv}.

Because the weights $\omega_i$ depend on the unknown distributions $F_i$ and $G$, 
we propose an iterative reweighted estimation algorithm (Algorithm~\ref{alg:estimation}). 
We refer to the resulting estimator $\hat{\btheta}$ as the censored two-stage least squares (cTSLS) estimator.

\begin{algorithm}[!ht]
\caption{Iterative Reweighted Estimation Procedure}
\label{alg:estimation}
\begin{algorithmic}[1]
  \State \textbf{Initialize:} Set $\omega_i \gets 1$ for all $i$. 
         Fit the unweighted least squares regression to obtain $\hat{\btheta}^{(0)}$. 
  \For{$k = 0, 1, 2, \dots$}
    \State Compute fitted values $\hat{Y}_i^{(k)}=
    \hat\beta_0^{(k)} + \hat\beta_1^{(k)} (\hat\alpha_0^{(k)} + \hat\balpha_1^{(k) \top} \bZ_i + \hat\balpha_2^{(k) \top}\bD_i) + \hat\bbeta_2^{(k)\top}\bD_i$.
    \State Compute the second stage right-censored residuals $r_i^{(k)} \gets Y_{i\hat{G}}^* - \hat{Y}_i^{(k)}$.
    \State Estimate $F_i$ by applying the Kaplan--Meier estimator 
           to $(r_i^{(k)}, \delta_i)$, and denote it by $\hat{F}_i^{(k)}$.
    \State Substitute $\hat{F}_i^{(k)}$ and $\hat{Y}_i^{(k)}$ 
           into \eqref{eqn:sv_var} to compute 
           $\widehat{\Var}(Y_{i\hat{G}}^*)$.
    \State Update weights: $\omega_i^{(k+1)} \gets \widehat{\Var}(Y_{i\hat{G}}^*)^{-1}$.
    \State Refit the weighted least squares regression with reweights $\omega_i^{(k+1)}$ 
           to obtain $\hat{\btheta}^{(k+1)}$.
    \If{$\|\hat{\btheta}^{(k+1)} - \hat{\btheta}^{(k)}\|_\infty < \texttt{tol}$ or $k>\texttt{kmax}$,}
       \State \textbf{stop}
    \EndIf
  \EndFor
  \State \textbf{Output:} Final estimate $\hat{\btheta}=\hat{\btheta}^{(k+1)}$.
\end{algorithmic}
\end{algorithm}
\noindent Note: In our empirical experience, the algorithm converges very quickly. 
With $\texttt{tol} = 10^{-3}$ and $\texttt{kmax} = 10$, convergence is typically achieved within 10 iterations.

In the next subsection, we show that the cTSLS estimator $\hat{\btheta}$ is asymptotically normal conditional on the weights, and we derive a consistent sandwich variance estimator.

\subsubsection{Asymptotic Property and Inference}\label{sec: Asymptotic Property}
Note that for a given set of weights $\{\omega_i\}$, the cTSLS estimator $\hat{\btheta}$ is the solution to the estimating equations
\begin{align*}
    S_n(\btheta; \hat{G}) = 0, 
\end{align*}
where the estimating function is
\begin{align}
    S_n(\btheta; \hat{G}) =
    \begin{bmatrix}
        \Psi_{1}(\btheta) \\
        \Psi_{2}(\btheta; \hat{G})
    \end{bmatrix}, \label{eq:score}
\end{align}
with the first- and second-stage components given by
\begin{align}
\Psi_{1}(\btheta) &= \sum_{i=1}^{n}
\begin{bmatrix}
1 \\
\bZ_{i} \\
\bD_i
\end{bmatrix}
\left(X_{i}-\mu_{X_i}\right), \label{eq:psi1} \\
\Psi_{2}(\btheta; \hat{G}) &= \sum_{i=1}^{n} \omega_i
\begin{bmatrix}
1 \\
\mu_{X_i} \\
\bD_i
\end{bmatrix}
\left(Y^*_{i\hat{G}}-\mu_{Y_i}\right), \label{eq:psi2} 
\end{align}
where
\begin{align}
\mu_{X_i} &= \E\left[X_i \mid \bZ_i,\bD_i\right] 
= \alpha_{0} + \balpha_{1}^{\top}\bZ_{i} + \balpha_{2}^{\top}\bD_i, \label{eqn:E(X)} \\
\mu_{Y_i} &= \E\left[Y_i \mid \bZ_i,\bD_i\right] 
= \beta_{0} + \beta_{1} \mu_{X_i} + \bbeta_{2}^{\top}\bD_i. \label{eqn:E(Y)}
\end{align}

Since the synthetic variables $Y_{i\hat G}^*$ are not independent across subjects, standard asymptotic theory for the GEE framework cannot be applied directly to derive the asymptotic properties of $\hat{\btheta}$. 
To address this issue, we extend the result of \citet{lai1995asymptotic} to our IV–AFT setting, as stated in the following Lemma \ref{lemma G_n}.

\begin{lemma}
Let $\btheta_0$ denote the true parameter. Under suitable regularity conditions, we have
\begin{align*}
    \frac{1}{\sqrt{n}}\left\{S_{n}(\btheta_0; \hat{G})-S_{n}(\btheta_0; G)\right\}=\frac{1}{\sqrt{n}}\sum_{j=1}^{n} 
    \left[\begin{array}{c}
\mathbf{0} \\
\Psi_{2j}^{*}\left(\btheta_{0}; G\right)
\end{array}\right]+ o_p(1),\quad \mbox{as $n\to \infty$},
\end{align*}
where $\Psi_{2j}^{*}(\btheta_0; G)$ is defined in \eqref{eq:psi2j} of Appendix B.
\label{lemma G_n}
\end{lemma}
The proof of Lemma \ref{lemma G_n} is given in Appendix B.

It follows from Lemma \ref{lemma G_n} that 
\begin{align*}
    \frac{1}{\sqrt{n}} S_{n}\left(\btheta_{0}; \hat{G}\right) & = \frac{1}{\sqrt{n}}S_{n}\left(\btheta_{0}; G\right) + \frac{1}{\sqrt{n}}\left\{S_{n}\left(\btheta_{0}; \hat{G}\right) - S_{n}\left(\btheta_{0}; G\right)\right\}\\
    & =\frac{1}{\sqrt{n}} \sum_{j=1}^{n}\left[\begin{array}{c}
\Psi_{1j}\left(\btheta_{0}; G\right) \\
\Psi_{2j}\left(\btheta_{0}; G\right)+\Psi_{2j}^{*}\left(\btheta_{0}; G\right)
\end{array}\right]+o_{p}(1),
\end{align*}
where $\Psi_{1i}$ and $\Psi_{2i}$, defined in \eqref{eq:psi1} and \eqref{eq:psi2}, denote the individual contributions to the estimating functions, and $\Psi_{2i}^{*}$ accounts for the additional variability due to estimation of the censoring distribution $G$.
This representation allows us to establish the asymptotic normality of $\frac{1}{\sqrt{n}} S_{n}\left(\btheta_{0}; \hat{G}\right)$ by applying the standard Central Limit Theorem. Combined with a Taylor series expansion, this yields the asymptotic distribution of the cTSLS estimator $\hat{\btheta}$, as stated in Theorem~\ref{theorem 1} below.


\begin{theorem}\label{theorem 1}
 Under suitable regularity conditions, conditional on the weight $\omega_i>0$, $\sqrt{n}\left(\hat{\btheta}-\btheta_0\right)$ has a limiting normal distribution with mean 0 and covariance matrix $\bA^{-1} \bB \bA^{-\top}$, where $\bA$ has the block form
\begin{align*}
    \bA &= \left[\begin{array}{ll}
\bA_{11} & \bA_{12} \\
\bA_{21} & \bA_{22}
\end{array}\right],
\end{align*}
with
\begin{align*}
\bA_{11} &= \E \left\{\left[\begin{array}{l}
1\\ \bZ_i \\ \bD_i
\end{array}\right] \left[1,\quad \bZ_i, \quad \bD_i\right]\right\},\\
\bA_{12} &= \mathbf{0},\\
\bA_{21} &= \E\left\{ \omega_i \left[\begin{array}{l}
\beta_1\\ \beta_1\mu_{X_i}+\mu_{Y_i}-Y_{iG}^*  \\ \beta_1\bD_i
\end{array}\right] \left[1,\quad \bZ_i ,\quad \bD_i\right]\right\},\\
\bA_{22} &=\E\left\{ \omega_i \left[\begin{array}{l}
1\\ \mu_{X_i}  \\ \bD_i
\end{array}\right]\left[1,\quad \mu_{X_i}, \quad \bD_i\right]\right\},
\end{align*}
and
\begin{align*}
    \bB=\E\left[\begin{array}{l}
\Psi_{1 j}({\btheta_0}) \\
\Psi_{2 j}({\btheta_0}; {G})+\Psi_{2 j}^{*}({\btheta_0}; {G})
\end{array}\right]\left[\begin{array}{l}
\Psi_{1 j}({\btheta_0}) \\
\Psi_{2 j}({\btheta_0}; {G})+\Psi_{2 j}^{*}({\btheta_0}; {G})
\end{array}\right]^{\top}.
\end{align*}
\end{theorem}

The proof of  Theorem \ref{theorem 1} is provided in the Appendix \ref{sec: proof theorm}. 

It is worth noting that Theorem~\ref{theorem 1} implies that, for any $k \ge 0$, the estimator $\hat\btheta^{(k)}$ is asymptotically normal, with a variance–covariance matrix that depends on the corresponding weights.

\subsubsection{Variance Estimation}
To obtain a consistent estimator of the asymptotic variance–covariance matrix $\bA^{-1}\bB\bA^{-\top}$ of $\hat{\btheta}$, we define the sample analogues \(\hat{\bA}\) and \(\hat{\bB}\) as follows:

\begin{align*}
        \hat{\bA}  = - \frac{S^{\prime}_{n}({\btheta}; \hat{G})}{n}\Bigg|_{\btheta=\hat{\btheta}}= - \left[\begin{array}{ll}
\frac{\partial \Psi_1({\btheta})}{n \partial \balpha^{\top}} & \frac{\partial \Psi_1({\btheta})}{n \partial \bbeta^{\top}} \\
\frac{\partial \Psi_2({\btheta}; \hat{G})}{n \partial \balpha^{\top}} & \frac{\partial \Psi_2({\btheta}; \hat{G})}{n \partial \bbeta^{\top}}
\end{array}\right]\Bigg|_{\btheta=\hat{\btheta}},
\end{align*}
where
\begin{align*}
    \frac{\partial \Psi_1({\btheta})}{n \partial \balpha^{\top}}=&\frac{1}{n} \sum_{i=1}^n\left[\begin{array}{l}
1\\ \bZ_i \\ \bD_i
\end{array}\right] \left[1,\quad \bZ_i, \quad \bD_i\right], \\
 \frac{\partial \Psi_1({\btheta})}{n \partial \bbeta^{\top}}=& \mathbf{0}, \\
 \frac{\partial \Psi_2({\btheta}; \hat{G})}{n \partial \balpha^{\top}} =&\frac{1}{n}\sum_{i=1}^n\omega_i \left[\begin{array}{l}
\beta_1\\ \beta_1\mu_{X_i}+\mu_{Y_i}-Y_{i\hat{G}}^*  \\ \beta_1\bD_i
\end{array}\right] \left[1,\quad \bZ_i ,\quad \bD_i\right],  \\
 \frac{\partial \Psi_2({\btheta}; \hat{G})}{n \partial \bbeta^{\top}}=&\frac{1}{n} \sum_{i=1}^n \omega_i \left[\begin{array}{l}
1\\ \mu_{X_i}  \\ \bD_i
\end{array}\right]\left[1,\quad \mu_{X_i}, \quad \bD_i\right].
\end{align*}

Additionally, define
\begin{align*}
    \hat{\bB}=\frac{1}{n} \sum_{j=1}^{n}\left\{\left[\begin{array}{l}
\Psi_{1 j}(\hat{\btheta}) \\
\Psi_{2 j}(\hat{\btheta}; \hat{G})+\widehat{\Psi}_{2 j}^{*}(\hat{\btheta}; \hat{G})
\end{array}\right]\left[\Psi_{1 j}^{T}(\hat{\btheta}), \quad \Psi_{2 j}^{T}(\hat{\btheta}; \hat{G})+\widehat{\Psi}_{2 j}^{* T}(\hat{\btheta}; \hat{G})\right]\right\},
\end{align*}
where
\begin{align*}
    \widehat{\Psi}_{2 j}^{*}(\hat\btheta; \hat{G}) &= \int_{s} \frac{\sum_{i=1}^{n} \omega_{i} \hat \bv_i \I_{\left\{s<\tilde{Y}_{i}\right\}} \cdot \int_{s}^{\tilde{Y}_{i}-} \frac{1}{\left\{1-\hat{G}_{n}(u)\right\}} d u}{Y(s)}   \left(d N_{j}(s)-Y_{j}(s) \cdot \frac{d N(s)}{Y(s)}\right),
\end{align*}
with the counting and at-risk processes $N_j(t)$ and $Y_j(t)$, as well as the aggregated processes $N(t)$ and $Y(t)$, defined in Appendix~\ref{sec: proof lemma} (Equations~\eqref{eq:counting} and \eqref{eq:countingall}),
and
\begin{align*}
    \hat \bv_i 
= \left[\begin{array}{l} 1 \\ \mu_{X_i} \\ \bD_i \end{array}\right]\Bigg|_{\btheta=\hat{\btheta}}.
\end{align*}

It follows, under suitable regularity conditions, that
\begin{align*}
    \hat{\bA} \;\xrightarrow{p}\; \bA 
    \quad \text{and} \quad 
    \hat{\bB} \;\xrightarrow{p}\; \bB, 
    \quad \text{as } n \to \infty.
\end{align*}
Consequently, this leads to a consistent sandwich variance estimator for $\btheta$ of the form
\[
\widehat{\Var}(\hat{\btheta}) \;=\; \hat{\bA}^{-1}  \hat{\bB}  \hat{\bA}^{-\top}.
\]

\section{Simulation study}\label{sec:sim_study}

To assess the finite sample operating characteristics of the proposed cTSLS method, we perform simulation experiments across a range of sample sizes and censoring proportions under several model settings.

\subsection{Data Generation}
For data generation we use the following reduced-form reparameterization, which is algebraically equivalent to \eqref{eq:X_i}–\eqref{eq:Y_i}:
\begin{align}
  X_i &= \balpha_1^\top \bZ_i + \balpha_2^\top \bD_i + \xi_{1i}, \label{eq:X_i_update}\\
  Y_i &= \beta_1 X_i + \bbeta_2^\top \bD_i + \xi_{2i}, \label{eq:Y_i_update}
\end{align}
where \(\bZ_i\in\mathbb R^2\) are instruments and \(\bD_i\in\mathbb R^2\) are observed confounders. 
Endogeneity is induced via the errors
\begin{align*}
    \xi_{1i} = \alpha_0 + \varepsilon_{1i},
  \qquad
  \xi_{2i} = \beta_0 + \varepsilon_{2i}.
\end{align*} 
To generate data from the above model, we first set the regression parameters in Equation \eqref{eq:X_i_update} as \( \balpha_1 = (0.5, 0.5)^\top \), and \( \balpha_2 = (0.3, 0.3)^\top \). Additionaly, we specify the regression parameters in Equation \eqref{eq:Y_i_update}  as \( \beta_1 = 1 \), along with \( \bbeta_2 = (0.5, 0.5)^\top \). Next, we simulate a two-dimensional instrument \( \bZ_i \) and a two-dimensional observed confounder \( \bD_i \), drawn from standard bivariate normal distributions \( \mathcal N(0, 0.8^2 * \bI_2) \) and \( \mathcal N(0, \bI_2) \) respectively.  We then generate the bivariate error distribution \( (\xi_{1i}, \xi_{2i})^\top \) based on two different scenarios, one is bivariate normal and another is a three–component mixture of bivariate normal, as detailed in Table \ref{tab:error_distribution}. Finally, for subject $i$, we calculate outcomes $X_i$ and uncensored log event times $Y_i$ based on Equation \eqref{eq:X_i_update} and \eqref{eq:Y_i_update}.

\begin{table}[ht]
\centering
\caption{Specification of the bivariate distribution of $(\xi_{1i}, \xi_{2i})^\top$ under two simulation scenarios.}
\label{tab:sim-settings}
\begin{tabular}{lcccccc}
\toprule
{Component} & {Mean$_1$} & {Mean$_2$} & {Var$_1$} & {Var$_2$} & {$\rho$} & {Prop.} \\
\midrule
\multicolumn{7}{l}{\textbf{(1) Single Gaussian}}\\
\addlinespace[2pt]
1 & 0.00 & 0.00 & 0.50 & 1.00 & -0.42 & 1.00 \\
\addlinespace[6pt]
\multicolumn{7}{l}{\textbf{(2) Three-component Gaussian mixture}}\\
\addlinespace[2pt]
1 & 5.00 & 4.00 & 0.20 & 1.00 &  \phantom{-}0.70 & 0.50 \\
2 & 5.00 & 1.00 & 0.40 & 0.50 &  \phantom{-}0.50 & 0.30 \\
3 & 5.00 & 5.00 & 0.30 & 2.00 &             -0.90 & 0.20 \\
\bottomrule
\end{tabular}\label{tab:error_distribution}
\end{table}

To incorporate censoring into the data, we generate censoring times according to a pre-specified mean censoring rate $\pi_c$. We assume that the censoring distribution is Normal and has the same standard deviation as the event time distribution. The mean of the censoring distribution is chosen to achieve the desired censoring proportion. Specifically, we proceed as follows:
\begin{enumerate}
\item Generate a large population ($N=100{,}000$) of log event times $Y_i^{\text{pop}}$ via above model and compute the standard deviation $\sigma_c$ of $Y_i^{\text{pop}}$.
\item Draw a temporary log censoring time $r_i\sim\mathcal N(0,\sigma_c^2)$. 
\item Define the censoring shift $\mu$ and set $ C_i = \mu + r_i$. Using the pre-specified censoring proportion $\pi_c$,  determine a censoring shift $\mu$ by solving for the appropriate threshold.
\end{enumerate}
After determining $\mu$ and $\sigma_c$, the sample data for each subject $i$ is generated as follows:
\begin{enumerate}
   \item Generate $Y_i$ from above model for log event times.
\item Draw $ C_i \sim \mathcal N(\mu,\sigma_c^2)$.
\item Define the observed time and event indicator as
\begin{align*}
    T_i=\exp(\tilde{Y}_i)=\exp\left\{\min(Y_i,C_i)\right\}, \qquad \delta_i=\I\{Y_i\le C_i\},
\end{align*}
where $T_i$ is the observed time and $\delta_i$ is the event indicator. This  process yields independent random censoring with an average censoring proportion close to $\pi_c$.
\end{enumerate}

We vary the sample size $n \in \{100, 500, 1000\}$ and censoring proportion $\pi_c \in \{0, 0.25, 0.5, 0.75\}$. 
Each scenario is replicated 500 times. Within each replicate, we fit the following models:  

\begin{enumerate}
    \item the proposed cTSLS estimator (using Leurgans’ synthetic outcomes with iterative weighting to account for heteroskedasticity in $Y_{i\hat G}^*$),
    \item an unweighted cTSLS estimator,
    \item the parametric Bayesian instrumental variable (PBIV) model \citep{li2015bayesian}, and
    \item a censored one-stage least squares (cOLS) estimator based on Leurgans’ synthetic outcomes without an instrumental variable.
\end{enumerate}

For the cTSLS estimators, we also compute sandwich standard errors. 
Performance is evaluated in terms of estimation bias (Figure~\ref{fig:bias}), variance estimation (Figure~\ref{fig:var leur}), 95\% coverage across replicates (Figure~\ref{fig:coverage}), and computation time (Figure~\ref{fig:running time}).

\subsection{Bias analysis}

\begin{figure}[ht]
\centering
\includegraphics[width=1\textwidth]{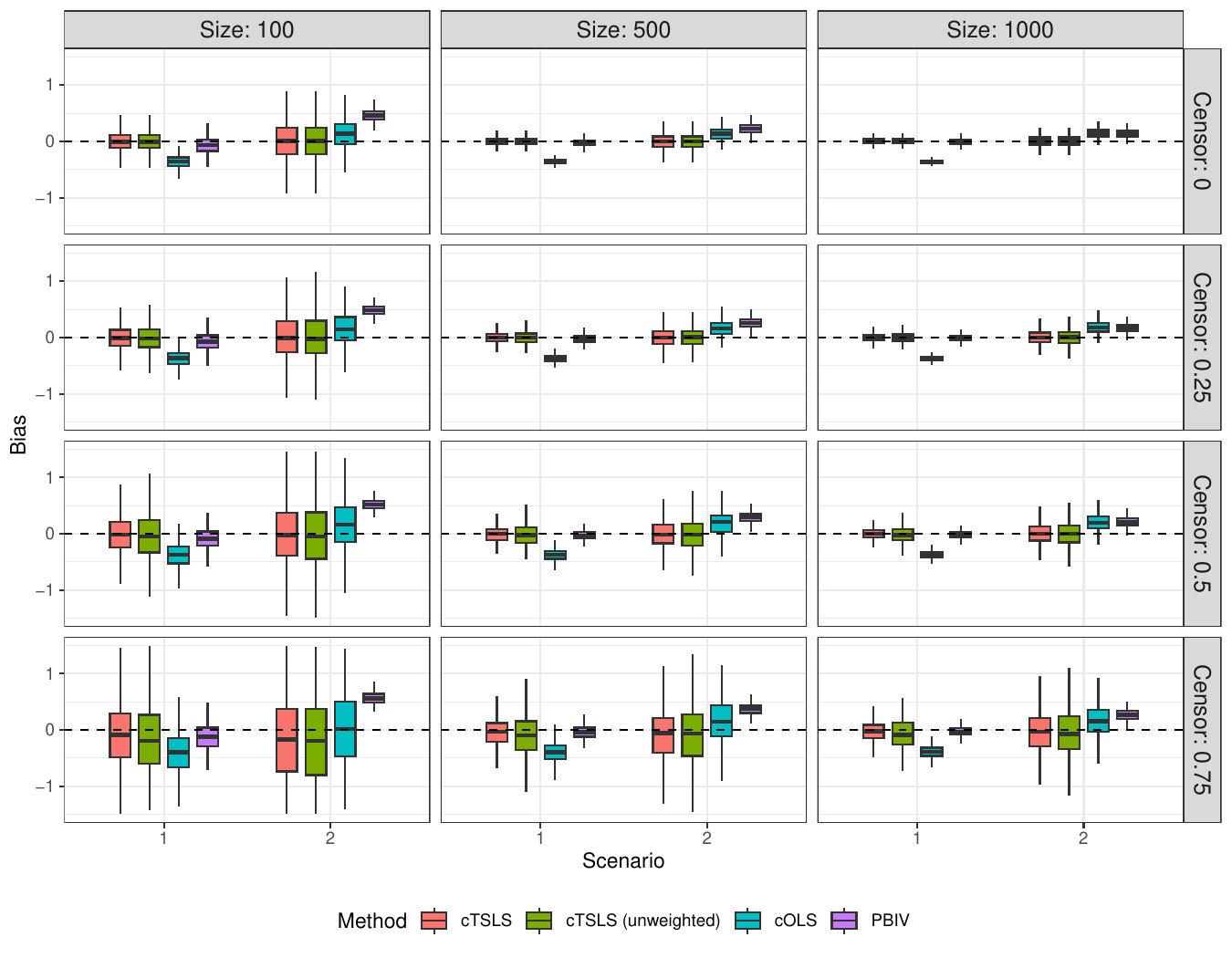}
\caption{Boxplots of estimation error based on 500 Monte Carlo replicates for four methods: cTSLS with subject-specific weights, unweighted cTSLS, censored one–stage least–squares estimator using Leurgans’ synthetic outcomes without instruments (cOLS) and parametric Bayesian instrumental variable model (PBIV). Columns correspond to sample sizes and rows to censoring rates. Within each panel, results are shown for two bivariate error designs (Scenario 1: Bivariate Gaussian, Scenario 2: Mixed Bivariate Gaussian).}
\label{fig:bias}
\end{figure}

Figure \ref{fig:bias} illustrates the biases of different methods across various sample sizes, censoring rates, and two error distributions (Bivariate Gaussian and Gaussian mixture). The weighted cTSLS estimator consistently shows smaller bias and variance than the unweighted cTSLS in all scenarios. Under Bivariate Gaussian errors, both cTSLS and PBIV are unbiased. However, under the Gaussian mixture, PBIV remains biased even at large \(n\), reflecting its sensitivity to non-Gaussian error structure. The cOLS estimator is biased across all settings because it ignores endogeneity. The Figure \ref{fig:bias} also illustrates the impact of sample size and censoring rate on estimator performance: Bias decreases as sample size increases, while higher censoring proportions are associated with larger bias and greater variability.

\subsection{Variance analysis}

\begin{figure}[ht]
\centering
\includegraphics[width=1\textwidth]{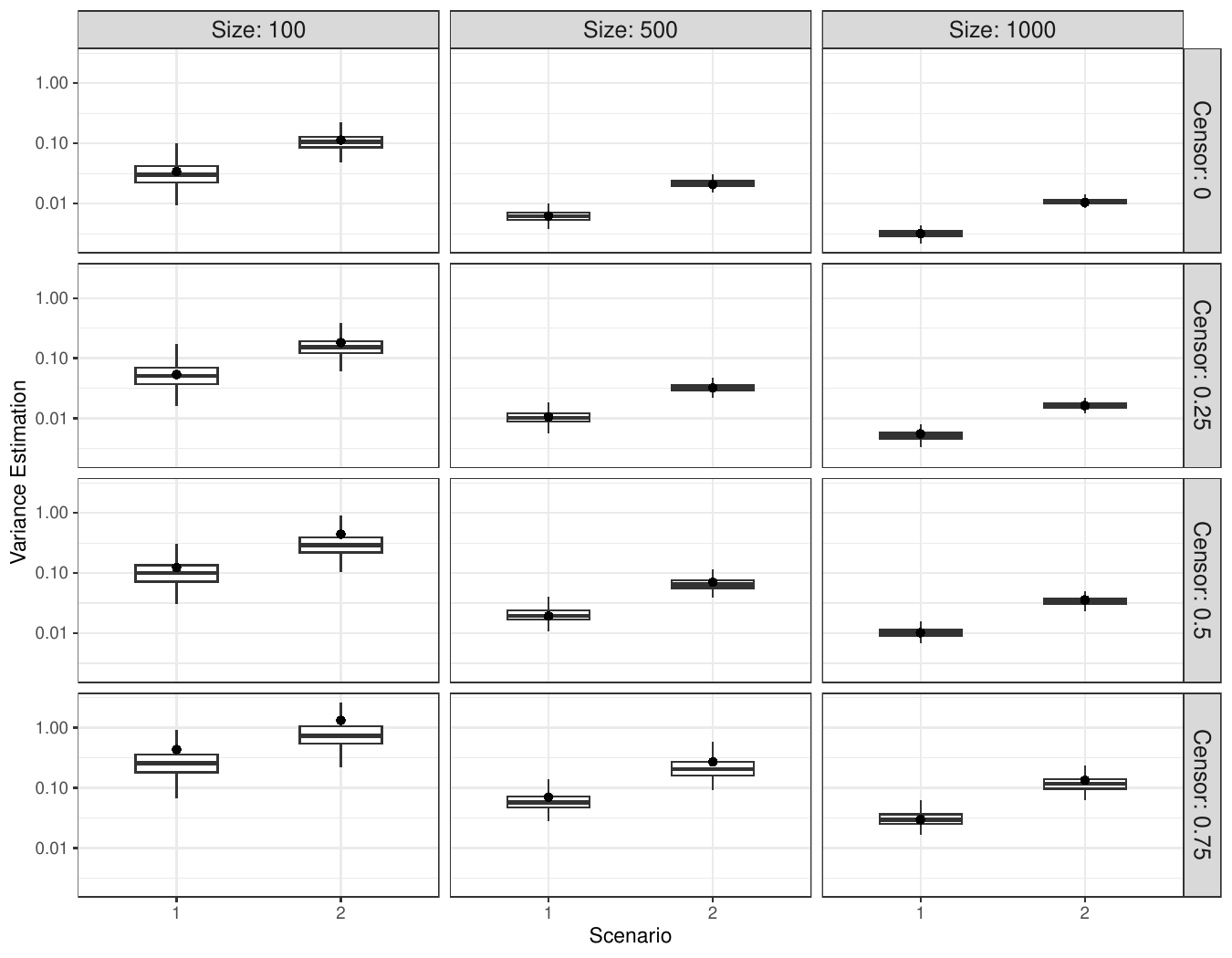}
\caption{Boxplots of sandwich variance estimates for the cTSLS estimator across 500 Monte Carlo replicates. The black dot indicates the variance of the causal effect calculated from the Monte Carlo samples, which is treated as the true variance. Columns correspond to sample sizes and rows to censoring rates. Within each panel, results are shown for two bivariate error designs (Scenario 1: Bivariate Gaussian, Scenario 2: Mixed Bivariate Gaussian).}
\label{fig:var leur}
\end{figure}


Figure \ref{fig:var leur} compares our sandwich variance estimates for the cTSLS estimator across sample sizes and censoring rates against the empirical Monte Carlo variance computed over 500 replicates (black dots). As censoring increases, the variance estimates become larger and show a modest deviation from the Monte Carlo benchmark, especially at small sample size $(n=100)$. As $n$ grows, the sandwich estimator approaches the benchmark closely, showing that our sandwich estimator is a consistent estimator.

\subsection{Coverage analysis}

\begin{figure}[ht]
\centering
\includegraphics[width=1\textwidth]{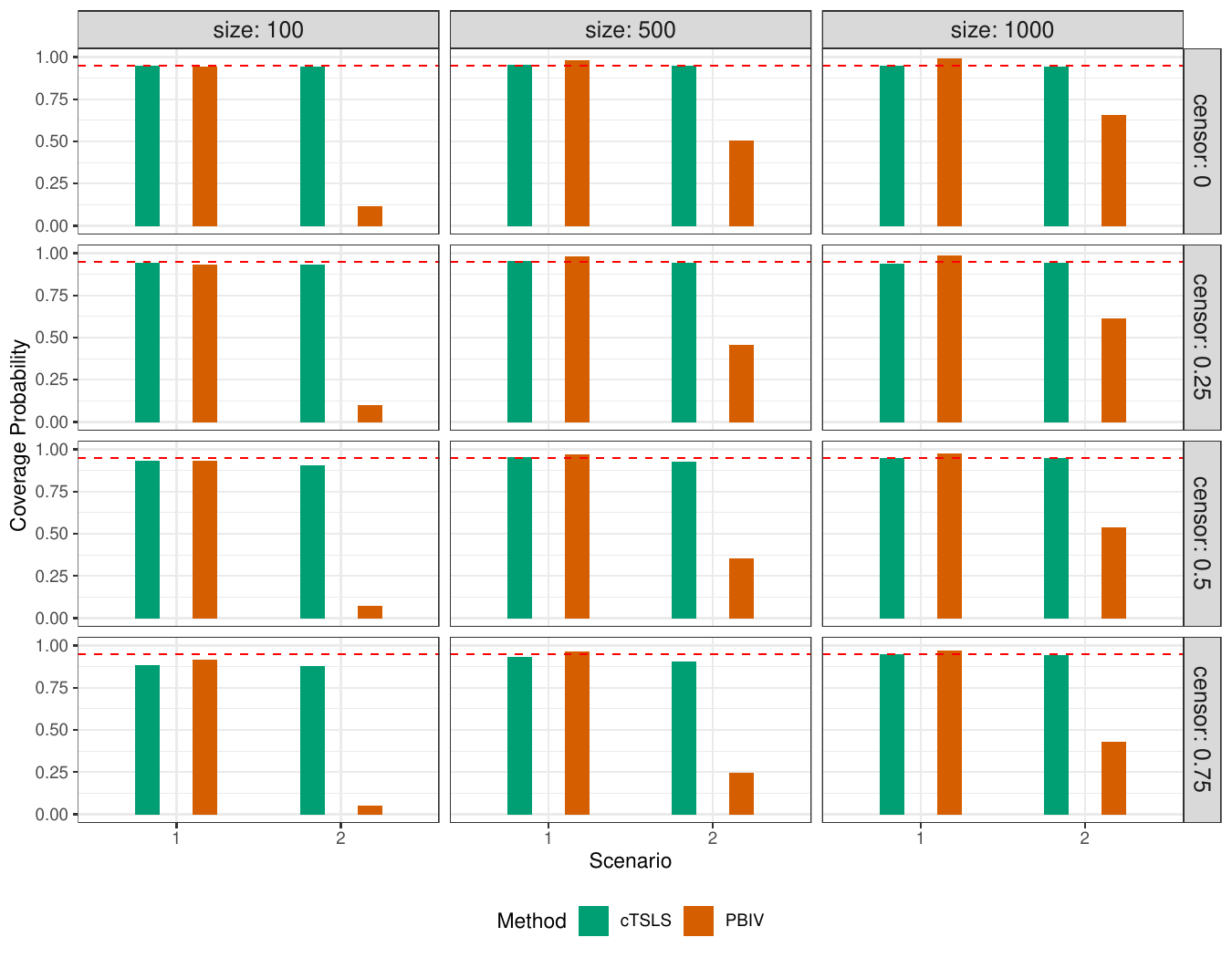}
\caption{Coverage probabilities (nominal 95\% CIs) from 500 Monte Carlo replicates for two models: (1) cTSLS with subject-specific weights and sandwich estimator, and (2) PBIV. Columns correspond to sample sizes and rows to censoring rates. Within each panel, results are shown for two bivariate error designs (Scenario 1: Bivariate Gaussian, Scenario 2: Mixed Bivariate Gaussian).}
\label{fig:coverage}
\end{figure}

Figure \ref{fig:coverage} shows the coverage probabilities for nominal 95\% confidence intervals for the cTSLS and PBIV estimators. Under the simple bivariate–Gaussian error design, both methods achieve near-nominal coverage. Under the Gaussian-mixture error design, the cTSLS method maintains coverage close to 95\% across sample sizes and censoring rates, whereas PBIV shows low coverage, especially in smaller samples, indicating sensitivity to violations of the bivariate normality assumption.

\subsection{Running time analysis}

\begin{figure}[ht]
\centering
\includegraphics[width=0.9\textwidth]{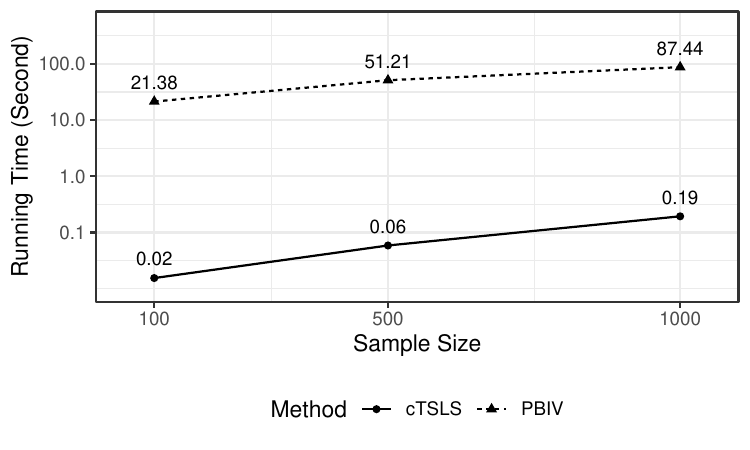}
\caption{Mean running time (seconds) for cTSLS (solid) and PBIV (dashed) across 500 Monte Carlo replicates. For each sample size, times are averaged over all error scenarios and censoring rates (which show similar costs at fixed $n$). Benchmark environment: Apple M2 (8-core), 16GB RAM; macOS 24.6; R~4.4.2.}
\label{fig:running time}
\end{figure}

Figure~\ref{fig:running time} compares the mean running time for cTSLS and PBIV across 500 Monte Carlo replicates. Averaging over error designs and censoring rates (which show similar costs at fixed $n$), cTSLS is consistently more than 300-fold faster than PBIV for sample sizes from $n=100$ to $n=1000$. This substantial speed advantage makes cTSLS especially practical for larger cohorts' analysis.

\section{Real Data Example: UK Biobank}\label{sec:real_study}

To illustrate the proposed cTSLS method, we consider data from the UK Biobank, a large prospective cohort study launched in 2006 to investigate genetic and environmental determinants of disease. The study initially enrolled approximately 500{,}000 participants aged 40–69 over a four-year recruitment period, with follow-up planned for at least 30 years. For this analysis, we used follow-up data through January 1, 2018. Participants were followed until this date, death, or loss to follow-up. The exposure variable $X$ is systolic blood pressure (SBP), and the outcome variable $Y$ is the logarithm of time to cardiovascular disease (CVD) from diagnosis of diabetes mellitus (DM). We identified 18{,}761 individuals diagnosed with DM prior to any CVD event, among whom 2{,}004 experienced CVD and 16{,}757 were censored. The median follow-up time was 12.31 years (IQR: 6.21, 18.00). The cohort includes 7{,}342 women (mean age at first assessment: 59.03, SD 7.26) and 11{,}419 men (mean age at first assessment: 59.37, SD 7.14).


We use the initial SBP measurement as the instrumental variable $\bZ$ for $X$ (first-stage $R^2 \approx 0.26$). 
The vector of observed confounders $\bD$ includes sex, age at first assessment, Townsend deprivation index, body mass index (BMI), physical activity (MET minutes), ISCED education level, genetic ancestry principal components (PC1–PC5), smoking status, and alcohol drinking status. 
All models adjust for $\bD$.  

We fit three models: (1) the proposed cTSLS estimator, (2) a censored one-stage least squares (cOLS) estimator based on Leurgan's synthetic outcomes without an instrumental variable, and (3) the parametric Bayesian IV (PBIV) model. 
Prior studies \citep{anderson2011blood,wan2020age,asia2007systolic} have suggested that higher SBP increases CVD risk. 
Accordingly, we expect a negative estimated causal effect under the AFT-type models and a positive estimated hazard ratio under the Cox model. 
Estimates of the causal effect are reported in Table~\ref{tab:ukb est}.

\begin{table}[ht]
\centering
\caption{Causal effect estimates from different methods for the UK Biobank data.}
\label{tab:ukb est}
\begin{tabular}{@{}lrrrrr@{}}
\toprule
Method  & $\hat\beta$ & SD & 95\% CI & $p$-value & Running time (s) \\ 
\midrule
cTSLS   & -1.97  & 0.63 & (-3.21, -0.73) & 0.002    & 22.33 \\
cOLS    & -1.01  & 0.37 & (-1.63, -0.31) & 0.006    & 21.99 \\
PBIV    &  1.03  & 0.20 & (0.62, 1.34)   & $<0.001$ & 33680.13 \\
\bottomrule
\end{tabular}
\end{table}

Table~\ref{tab:ukb est} summarizes the SBP effect estimates from the three approaches. 
The cTSLS estimator yields a negative causal effect ($\hat\beta = -1.97$, $p = 0.002$), consistent with the expectation that higher SBP shortens time to CVD. 
The cOLS estimator also produces a negative effect ($\hat\beta = -1.01$, $p = 0.006$), but by ignoring endogeneity it attenuates the magnitude of the effect toward zero, a pattern consistent with our simulation findings (see Figure~\ref{fig:bias}). 
In contrast, the PBIV model yields a positive effect ($\hat\beta = 1.03$, $p < 0.001$), which is counterintuitive given prior epidemiologic evidence that elevated SBP increases CVD risk. 
Moreover, the proposed cTSLS method demonstrated an impressive computational advantage over the PBIV method (22 seconds versus 33,680 seconds, i.e., more than 9 hours) for this large dataset.

\begin{figure}
    \centering
    \includegraphics[width=1\linewidth]{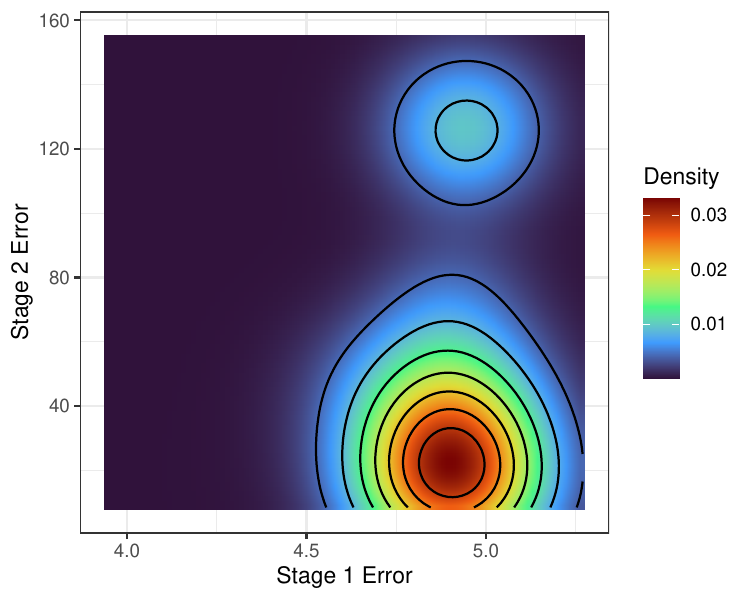}
    \caption{Kernel-smoothed heat map of the joint density of the stage-1 and stage-2 errors from cTSLS model. Color shows density magnitude. Black contour lines are isodensity curves at evenly spaced levels.}
    \label{fig:ukb contour}
\end{figure}

To investigate why the PBIV method yields a counterintuitive result, we examined its underlying assumption of bivariate normal errors. 
Specifically, based on the cTSLS estimates from the semiparametric IV–AFT model, we estimated the joint bivariate CDF of the two stage-specific errors $(\xi_1, \xi_2)$ using the nonparametric estimation method of \citet{akritas1994nearest}, which accommodates the setting where the first-stage residuals are uncensored while the second-stage residuals are subject to right censoring.
We then obtained a bivariate density estimate via kernel smoothing using the direct plug-in bandwidth selection method \citep{sheather1991reliable, wand1994kernel}. 
The contour plot of the estimated bivariate error density (Figure~\ref{fig:ukb contour}) clearly departs from normality, suggesting a mixture of multiple bivariate normal components. 
This observation may help explain the bias observed in the PBIV method and appears consistent with our simulation findings in Scenario~2.

\section{Discussion}\label{sec:discussion}

We have developed a censored two-stage least squares (cTSLS) estimator for instrumental variable analysis with right-censored outcomes under the semiparametric AFT framework. 
By incorporating synthetic variables and an iterative reweighting scheme, the proposed estimator extends the classical TSLS procedure to censored data while improving efficiency relative to naive approaches. 
Our theoretical results establish consistency and asymptotic normality, and simulation studies demonstrate favorable finite-sample performance compared to both unweighted estimators and fully parametric alternatives. 
The real-data analysis using the UK Biobank illustrates the practical utility of the method and highlights its advantages in terms of both interpretability and computational scalability.

Despite these strengths, several limitations merit discussion. 
First, the cTSLS estimator relies on the synthetic variable construction, which introduces additional variability relative to the fully observed case. 
Although our iterative reweighting scheme reduces inefficiency, finite-sample performance can deteriorate under heavy censoring. 
In particular, when the proportion of censored observations is high, the estimation of the censoring distribution via the Kaplan–Meier estimator becomes unstable, which in turn leads to greater variability in the synthetic outcomes and the resulting cTSLS estimates. 
In such settings, sandwich variance estimators may underestimate true variability, and inference should be interpreted with caution. 
A promising direction for future research is to develop IV methods targeting restricted mean survival time (RMST), which is often more robust under heavy censoring and may provide a stable alternative causal estimand.

Second, as with all IV methods, the validity of the cTSLS estimator depends on the strength and appropriateness of the chosen instrument. 
Weak instruments may lead to finite-sample bias and inflated standard errors, while violation of the exclusion restriction would compromise causal interpretability. 
Although our application to the UK Biobank highlights the utility of baseline SBP as an instrument, careful instrument selection and sensitivity analysis remain essential in practice.

Finally, while the cTSLS method is computationally efficient relative to parametric Bayesian alternatives, its current formulation assumes independent and identically distributed observations. 
Extensions to clustered or longitudinal settings, as well as adaptation to high-dimensional covariates, represent promising directions for future work.

In summary, the proposed cTSLS estimator enriches the toolkit for IV analysis with censored time-to-event outcomes, offering a semiparametric, computationally efficient, and theoretically justified alternative to existing approaches. 
At the same time, its limitations under heavy censoring underscore the importance of ongoing methodological development and cautious interpretation in challenging data environments.

\section{Software}

The proposed methods are implemented in the \texttt{cTSLS} package, available at  
\url{https://github.com/Larryzza/cTSLS}.

\section{Funding}\label{funding}

This work was supported by National Institutes of Health under grants (P30 CA-16042 (GL), UL1TR000124-02 (GL), R35 GM141798 (HZ), R01 HG006139 (HZ and JZ), and R01 DK142026 (HZ, JZ and GL); National Science Foundation under grants DMS 2054253 (HZ and JZ) and IIS 2205441 (HZ, JZ and GL).

\newpage
\bibliographystyle{plainnat} 
\bibliography{sn-bibliography} 

\clearpage    
\newpage
\pagenumbering{arabic} 
\setcounter{page}{1}

\section*{Appendices}
\begin{appendices}

\section{Derivation of Variance of Synthetic Variables}\label{section:var_sv}

In our cTSLS method, we adopt the Leurgans synthetic variable. As shown in \eqref{eq: sv leurgans}, we have,
\begin{align*}
    Y_{i{G}}^* = \tilde Y_i + \int_{-\infty}^{\tilde Y_i} \frac{G(t-)}{1 - G(t-)}  dt.
\end{align*}
Let $\Phi(u) = u + \int_{-\infty}^{u} \frac{G(t-)}{1 - G(t-)}  dt$. Then Leurgans synthetic variable can be written as
\begin{align}
Y_{iG}^* = \delta_i \Phi(Y_i) + (1-\delta_i) \Phi(C_i).\label{eq: leurgans2}
\end{align}
Under independent censoring, this construction satisfies
$\E(Y_{i{G}}^*\mid Y_i)=Y_i$, and $\E(Y_{i{G}}^*)=\E(Y_i)$. Therefore, 
\begin{align}
\V (Y_{iG}^*) &= \E (Y_{iG}^* - \E (Y_{i}))^2\nonumber\\ &= \E (Y_{i}^* - Y_{i} + Y_{i} - \E (Y_{i}))^2\nonumber\\
& = \Var(Y_i) + \E( (Y_{iG}^*)^2 ) - \E(Y_i^2). \label{eq: var decomp}
\end{align}
Since $\delta_i^2=\delta_i$ and $\delta_i(1-\delta_i)=0$, based on \eqref{eq: leurgans2}, we have,
\begin{align*}
    (Y_{iG}^*)^2 = \delta_i\Phi(Y_i)^2 + (1-\delta_i)\Phi(C_i)^2.
\end{align*}
Let $F_{Y_i}$ and $G$ denote the distribution of $Y_i$ and $C_i$. With $Y_i\perp C_i$,
\begin{align}
\E\{(Y_{iG}^*)^2\}
&= \E[ \I\{Y_i\le C_i\}\Phi(Y_i)^2 ]
   + \E[ \I\{C_i< Y_i\}\Phi(C_i)^2 ]\nonumber \\
&= \int \{1-G(s)\}\Phi(s)^2dF_{Y_i}(s) + \int \{1-F_{Y_i}(s)\}\Phi(s)^2dG(s). \label{Eq:Ey*^2}
\end{align}
It can be verified, after some algebra, that
\begin{align*}
\Var(Y_{iG}^*) 
= \Var(Y_i) 
  + 2\int_{G^{-1}(0)}^\infty \big\{1 - F_{Y_i}(s)\big\} \,\big\{\Phi(s)-s\big\}\, ds. 
\end{align*}

Since 
\[
\Phi(u) = u + \int_{-\infty}^{u} \frac{G(t-)}{1 - G(t-)}\, dt,
\qquad \text{with } \frac{G(t-)}{1 - G(t-)} = 0 \ \text{for } t < G^{-1}(0),
\]
we can equivalently write
\begin{align*}
\Var(Y_{iG}^*)
= \Var(Y_i) 
  + 2 \int_{G^{-1}(0)}^{\infty} \big\{1-F_{Y_i}(s)\big\}
    \int_{-\infty}^{s} \frac{G(t)}{1-G(t)}\, dt \, ds.
\end{align*}

Finally, using the decomposition $Y_i = \mu_{Y_i} + \varepsilon_{2i}$ with $F_i(s) = \Pr(\varepsilon_{2i} \le s)$, the upper limit of the inner integral shifts by $\mu_{Y_i}$:
\begin{align*}
\Var(Y_{iG}^*) 
= \Var(Y_i) 
  + 2 \int_{G^{-1}(0)}^\infty \big\{1 - F_i(s)\big\} 
    \int_{-\infty}^{\mu_{Y_i}+s} \frac{G(t)}{1-G(t)}\, dt \, ds.
\end{align*}

Here, we require $F_i^{-1}(1) < G^{-1}(1)$; otherwise, the integral diverges to infinity.

\section{Proof of Lemma 1}\label{sec: proof lemma}

Let 
\begin{align*} 
    S_n(\btheta; \hat{G})=\left[\begin{array}{l}\Psi_{1}(\btheta) \\ \Psi_{2}(\btheta; \hat{G})\end{array}\right]
\end{align*}
be the estimating function defined by \eqref{eq:score}, with $\Psi_{1}(\btheta)$ and
$\Psi_{2}(\btheta; \hat{G})$ given by \eqref{eq:psi1}
and \eqref{eq:psi2}, respectively.
Then
\begin{align} \label{eq:B2}
    S_{n}(\btheta_0 ; \hat{G})-S_{n}(\btheta_0 ; G) = \left[\begin{array}{l}\mathbf{0} \\ \Psi_{2}(\btheta_0; \hat{G})-\Psi_{2}(\btheta_0; G)\end{array}\right]. 
\end{align}
It is easy to see that
\begin{align} \label{eq:psi_dif}
    \Psi_{2}(\btheta_0; \hat{G})-\Psi_{2}(\btheta_0; G)=\sum_{i=1}^{n} \omega_{i} \bv_i \int_{-\infty}^{\tilde{Y}_{i}-} \frac{\hat{G}_{n}(t)-G(t)}{\{1-G(t)\}\left\{1-\hat{G}_{n} \left(t\right)\right\}} dt,
\end{align}
where
\begin{align*}
\bv_i 
=\left[\begin{array}{l} 1 \\ \mu_{X_i}\\ \bD_i\end{array}\right]\bigg|_{\btheta = \btheta_0}. 
\end{align*}

 For $j=1,...,n$, let 
 \begin{align} \label{eq:counting}
 N_j(t)=I(\tilde{Y}_j\le t, \delta_j=0), \quad Y_j(t)=I(\tilde{Y}_j\ge t) 
 \end{align}
be the counting process for the censoring event and at-risk process for subject $j$. Additionally,
let
 \begin{align} \label{eq:countingall}
 N(t)=\sum_{j=1}^n N_j(t), \quad Y(t)=\sum_{j=1}^n Y_j(t), 
 \end{align}
 be the aggregated counting process for the censoring event and the aggregated at-risk process, respectively.
Then,
\begin{align}
M_j(t) = N_j(t) - \int_0^t Y_j(s) d\Lambda_G(s), \quad j=1,\ldots,n, \label{eq:M}
\end{align}
are orthogonal local square integrable martingales, where
$$
\Lambda_G(s) \;=\; \int_{0}^{s} \frac{dG(u)}{1-G(u-)},
$$
is the cumulative hazard function of $C_i$.

According to \citet{gill1983large}, we have
\begin{align}
    \hat{G}_{n}(t) - G(t) 
    &= (1-G(t)) \int_{-\infty}^{t-} 
       \frac{1-\hat{G}_{n}(s)}{1-G(s)} 
       \frac{d \sum_{j=1}^{n} M_j(s)}{Y(s)}, 
       \label{eq:gill} \\[0.75em]
    \frac{1-\hat{G}_{n}(s-)}{1-G(s)} 
    &= 1+o_p(1). \label{eq:gill2}
\end{align}

Substituting \eqref{eq:gill} into \eqref{eq:psi_dif} yields
\begin{align*}
    \Psi_{2}(\btheta_0; \hat{G})-\Psi_{2}(\btheta_0; G)
    &= \sum_{i=1}^{n} \omega_{i} \bv_i
       \int_{-\infty}^{\tilde{Y}_{i}-} 
       \frac{1}{1-\hat{G}_{n}(t)}
       \int_{-\infty}^{t-} 
       \frac{1-\hat{G}_{n}(s)}{1-G(s)} 
       \frac{d \sum_{j=1}^{n} M_{j}(s)}{Y(s)} dt \\
    &= \sum_{j=1}^{n} \sum_{i=1}^{n} \omega_{i} \bv_i
       \int_{-\infty}^{\tilde{Y}_{i}-} 
       \frac{1-\hat{G}_{n}(s)}{(1-G(s))Y(s)} 
       \int_{s}^{\tilde{Y}_{i}-} 
       \frac{1}{1-\hat{G}_{n}(t)} dt \, dM_{j}(s).
\end{align*}

Applying \eqref{eq:gill2}, we obtain
\begin{align}
    \Psi_{2}(\btheta_0; \hat{G})-\Psi_{2}(\btheta_0; G) 
    &= \sum_{j=1}^{n} \int_{s} 
       \Bigg\{
       \frac{1}{Y(s)/n} \cdot \frac{1}{n}\sum_{i=1}^{n} 
       \omega_{i} \bv_i 
       \I_{\{s<\tilde{Y}_{i}\}} 
       \int_{s}^{\tilde{Y}_{i}-} 
       \frac{1}{1-G(t)} dt 
       \Bigg\} dM_{j}(s) + o_p(1) \nonumber \\
    &= \sum_{j=1}^{n} \int_{s} H(s)\, dM_{j}(s) + o_p(1), 
    \label{eq:x}
\end{align}
where the last step follows from the Law of Large Numbers applied to the integrand inside the braces, and $H(s)$ is the resulting limit given by
\begin{align}
    H(s) = \frac{1}{P(Y_i \geq s)P(C_i \geq s)} 
    \E \Bigg[ 
       \omega_{i} \bv_i \, \I_{\{s<\tilde{Y}_{i}\}} 
       \int_{s}^{\tilde{Y}_{i}-} \frac{1}{1-G(t)} dt 
    \Bigg].
    \label{eq:H}
\end{align}

Combining \eqref{eq:B2} and \eqref{eq:x} proves Lemma 1, where
\begin{equation} \label{eq:psi2j}
    \Psi_{2j}^{*}(\btheta, G) = \int_{s}  H(s) d M_{j}(s), \quad j=1,...,n,
\end{equation}
with $H(s)$ and $M_j(s)$ defined in \eqref{eq:H} and
\eqref{eq:M}, respectively.


\section{Proof of Theorem \ref{theorem 1}}\label{sec: proof theorm}

Firstly, by Lemma~\ref{lemma G_n}, we have
\begin{align*}
    \frac{1}{\sqrt{n}} S_{n}(\btheta_{0}; \hat{G}) 
    &= \frac{1}{\sqrt{n}} \Big\{ 
        S_{n}(\btheta_{0}; G) 
        + \big(S_{n}(\btheta_{0}; \hat{G}) - S_{n}(\btheta_{0}; G)\big) 
       \Big\} \\[0.5em]
    &= \frac{1}{\sqrt{n}} \sum_{j=1}^{n}
       \begin{bmatrix}
         \Psi_{1j}(\btheta_{0}) \\
         \Psi_{2j}(\btheta_{0}; G) + \Psi_{2j}^{*}(\btheta_{0}; G)
       \end{bmatrix} 
       + o_p(1).
\end{align*}

Note that $\E\{\Psi_{1j}(\btheta_{0})\}=0$, 
$\E\{\Psi_{2j}(\btheta_{0}; G)\}=0$, 
and $\E\{\Psi_{2j}^{*}(\btheta_{0}; G)\}=0$. 
Since $\Psi_{1j}(\btheta_{0})$, $\Psi_{2j}(\btheta_{0}; G)$, and $\Psi_{2j}^{*}(\btheta_{0}; G)$ are i.i.d., the Central Limit Theorem and Slutsky’s theorem imply
\begin{align}
    \frac{1}{\sqrt{n}} \sum_{j=1}^{n}
    \begin{bmatrix}
        \Psi_{1j}(\btheta_{0}) \\
        \Psi_{2j}(\btheta_{0}; G) + \Psi_{2j}^{*}(\btheta_{0}; G)
    \end{bmatrix} 
    \;\xrightarrow{d}\; \mathcal{N}(0, \bB), \label{eq:Sn_dis}
\end{align}
where
\begin{align*}
\bB &= \Var \!\left(
      \begin{bmatrix}
         \Psi_{1j}(\btheta_{0}) \\
         \Psi_{2j}(\btheta_{0}; G) + \Psi_{2j}^{*}(\btheta_{0}; G)
      \end{bmatrix}
      \right) \\
    &= \E \!\left\{
      \begin{bmatrix}
         \Psi_{1j}(\btheta_{0}) \\
         \Psi_{2j}(\btheta_{0}; G) + \Psi_{2j}^{*}(\btheta_{0}; G)
      \end{bmatrix}
      \begin{bmatrix}
         \Psi_{1j}(\btheta_{0}) \\
         \Psi_{2j}(\btheta_{0}; G) + \Psi_{2j}^{*}(\btheta_{0}; G)
      \end{bmatrix}^{\!\top}
      \right\}.
\end{align*}

 Since $S_n(\hat{\btheta};\hat G)=0$, there exists $\btheta^*$ between $\hat\btheta$ and $\btheta_0$ such that
\begin{align*}
    0 &= S_n(\btheta_{0}; \hat{G})
       + \Big\{ S_{n}^{\prime}(\btheta; \hat{G})\big|_{\btheta = \btheta^*} \Big\}
         (\hat{\btheta}-\btheta_{0}).    \end{align*}
Hence
        \begin{align} \label{eq:taylor}
    \sqrt{n}(\hat{\btheta}-\btheta_{0})
       &= \left\{ -\frac{S_{n}^{\prime}(\btheta^{*}; \hat{G})}{n} \right\}^{-1} 
          \frac{S_{n}(\btheta_{0}; \hat{G})}{\sqrt{n}}.
\end{align}

Under suitable regularity conditions, it can be shown that
\begin{align} \label{eq:jacobian}
    \left\|
    \frac{S_n^{\prime}(\btheta; \hat{G})}{n}\bigg|_{\btheta = \btheta^*}
    - \frac{S_n^{\prime}(\btheta; G)}{n}\bigg|_{\btheta = \btheta_0}
    \right\|
    \;\xrightarrow{p}\; 0,
\end{align}
where
\begin{align*}
    \frac{S_n^{\prime}(\btheta; G)}{n}\bigg|_{\btheta = \btheta_0} 
    &= \begin{bmatrix}
         \dfrac{\partial \Psi_1(\btheta)}{n \partial \balpha^{\top}} 
         & \dfrac{\partial \Psi_1(\btheta)}{n \partial \bbeta^{\top}} \\[1em]
         \dfrac{\partial \Psi_2(\btheta; G)}{n \partial \balpha^{\top}} 
         & \dfrac{\partial \Psi_2(\btheta; G)}{n \partial \bbeta^{\top}}
       \end{bmatrix}_{\btheta = \btheta_0}.
\end{align*}

Each block is an average of i.i.d. terms, so
\begin{align} \label{eq:A}
    -\frac{S_n^{\prime}(\btheta; G)}{n}\bigg|_{\btheta = \btheta_0}
    \;\xrightarrow{p}\; \bA
    = \begin{bmatrix}
        \bA_{11} & \bA_{12} \\
        \bA_{21} & \bA_{22}
      \end{bmatrix},
\end{align}
with
\begin{align*}
\bA_{11} &= \E \!\left\{
    \begin{bmatrix}
       1 \\ \bZ_i \\ \bD_i
    \end{bmatrix}
    \!\!
    \begin{bmatrix}
       1 & \bZ_i & \bD_i
    \end{bmatrix}
    \right\}, \\
\bA_{12} &= \mathbf{0}, \\
\bA_{21} &= \E \!\left\{ 
    \omega_i
    \begin{bmatrix}
       \beta_1 \\
       \beta_1\mu_{X_i}+\mu_{Y_i}-Y_{iG}^* \\
       \beta_1\bD_i
    \end{bmatrix}
    \!\!
    \begin{bmatrix}
       1 & \bZ_i & \bD_i
    \end{bmatrix}
    \right\}, \\
\bA_{22} &= \E \!\left\{ 
    \omega_i
    \begin{bmatrix}
       1 \\ \mu_{X_i} \\ \bD_i
    \end{bmatrix}
    \!\!
    \begin{bmatrix}
       1 & \mu_{X_i} & \bD_i
    \end{bmatrix}
    \right\}.
\end{align*}

Finally, combining \eqref{eq:Sn_dis}-\eqref{eq:A} with Slutsky’s theorem yields
\[
\sqrt{n}(\hat\btheta-\btheta_0)
\;\xrightarrow{d}\; \mathcal{N}\!\big(0,\; \bA^{-1} \bB \bA^{-\top}\big),
\]
with $\bA$ and $\bB$ defined as above.




\end{appendices}


\end{document}